# Origin and the role of device physics in the magnetic field effect in organic semiconductor devices

B. K. Li[1], H. T. He[1], W. J. Chen[1], M. K. Lam[2], K. W. Cheah[2], and J. N. Wang[1*]

[1]Department of Physics, Hong Kong University of Science and Technology, Clear Water Bay, Kowloon, Hong Kong, China
[2]Center of Advanced Luminance Materials, Department of Physics, Hong Kong Baptist University, Kowloon Tong, Hong Kong, China
*email: phjwang@ust.hk

**A small magnetic field (~30 mT) can effectively modulate the electroluminescence, conductance and/or photocurrent of organic semiconductor based devices, up to 10% at room temperature. This organic magnetic field effect (OMFE) is one of the most unusual phenomena of both organic electronics and, more basically, magnetism, since all device components are nonmagnetic. However, in spite of latest surge of research interest, its underlying mechanism is still hotly debated. Here we experimentally identify that the magnetic field induced increase of intersystem crossing rate (between either excitons or polaron pairs), and decrease of triplet exciton-polaron quenching rate are responsible for the observed OMFEs. The diversity of observed OMFE results, such as sign change and operating condition dependence, originates from the difference of devices physics.**

Organic semiconductor (OSE) based devices, such as organic light emitting diodes (OLEDs), organic photovoltaic cells and thin film transistors, as well as spintronic devices, have been successfully developed in the past few decades (ref. 1 and 2, and references therein). Although the magnetic field effect on luminescence, photoconductance in organic materials have been studied in the past [3,4], renewed research interest has surged very recently due to the recent discovery that a small magnetic field of ~ 30 mT can effectively modulate the electroluminescence (EL) and/or conductance up to ~10% at room temperature in organic devices, such as OLEDs, organic single layer sandwiched structures and organic photovoltaic devices, without any magnetic materials [5-19]. These organic magnetic field effects (OMFEs) do not possess magnetic field direction dependence. Obviously, the OMFE can be easily used as an external way to further increase the efficiency of organic photovoltaic devices, and to modulate the efficiency of OLEDs. New kinds of devices, such as magnetic sensors [19,21] and OLED based touch-screen devices using the OMFE can also be developed, further enriching the field of OSE applications.

OMFE is a universal phenomenon and has been observed and studied in various polymer and/or small molecular OSEs based devices. However, the observed experiment results were inconsistent and difficult to interpret. Firstly, the sign of OMFE can be positive or negative, depending on material [7], device structure such as layer thickness [14], and also on operating conditions such as applied bias and temperature [7,15]. Secondly, in study of OMFEs, usually only a low field (~30 mT) component, which saturates relatively fast, can be observed, while a high field (>100 mT) component has also been reported by several groups very recently [13,15,16]. Different physical processes, which are believed to be magnetic field dependent, have been proposed as the origin of OMFE, such as intersystem crossing (ISC) between polaron pairs (PPs) [5,8,9,11,14,20], triplet-triplet annihilation [12,13], polaron scattering by triplet excitons [14], triplet exciton quenching induced by polaron [14,31], influence on charge transfer states ($\Delta g$ mechanism) in blend structure [15], polaron pair formation [16], mobility change of minority carriers [17], and bipolaron formation [18]. From these physical processes, two kinds of models have been proposed to explain the complexity and divergence of experiment results. One is multi-process based model [14,15], in which it is proposed that different signs and/or different field dependences originate from different physical processes. The other is single-process based model [9,11,17], in which the sign change is a result of secondary effect. For example, the change of singlet/triplet exciton ratio, originating from the magnetic field induced reduction in ISC between PPs, can modulate the device current, either because of the different roles that singlet and triplet excitons plays in conductance [9], or through the change of recombination current [11]. The multi-process models are individual device dependent, while the single-process models do not explain the existence of different magnetic field dependent components. A simple universal model, which can explain not only the sign changes complexity but also the existence of different field dependent components, is lacking.

In this study, we show that magnetic field induced increase in singlet-triplet ISC rate (between either excitons or polaron pairs) and decrease in triplet exciton-polaron quenching rate are responsible for the observed OMFE. Based on these two processes, we show that the variety of observed phenomena, such as the coexistence of low- and high- field components, sign changes, as well as the material dependence, organic layer thickness dependence and operating conditions dependence, of the measured OMFE curves, is result of the change of devices physics.

In this work we used two small molecule OSEs,

N,N'-di(naphthalene-1-yl)-N,N'- diphenyl-benzidine (NPB) and tris-(8-hydroxyquinolinato) aluminium (Alq$_3$), both widely used in OLEDs as hole transport layer and electron transport/emission layer, respectively [22], as active layer in our single layer sandwiched devices, with a structure of ITO/OSE/cathode. One advantage of small molecular over polymer is the higher purity, which can exclude any extrinsic influence of the observed OMFEs [1]. Magnetic field effects on photocurrent (*MPC*) of the devices were systematically studied as a function of photon excitation energy ($E_{hv}$), applied bias (*V*), and layer thickness (*d*). The *MPC* is defined as:

$$MPC\% = \frac{PC(B) - PC(B=0)}{PC(B=0)} \times 100\% \quad (1)$$

where *PC*, i.e. photocurrent is the illumination induced device current change.

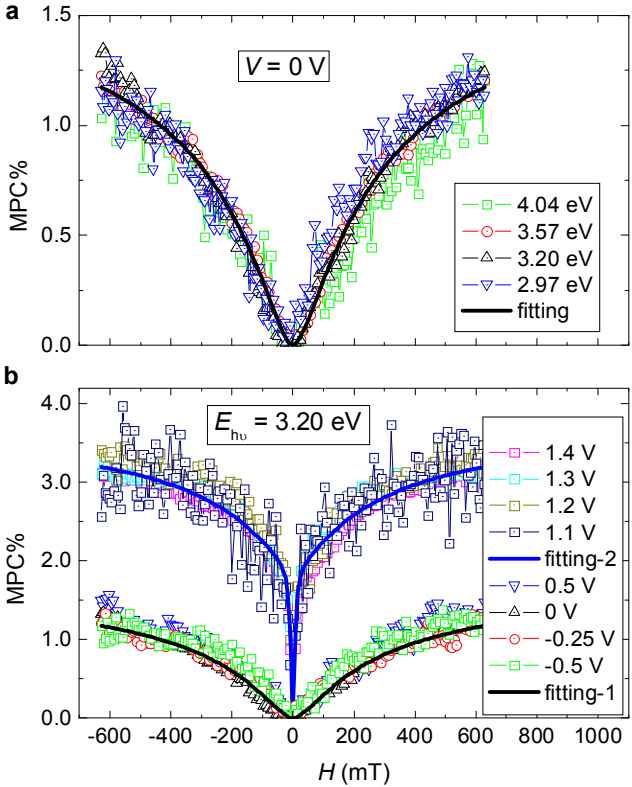

**Figure 1 MPC of device ITO/NPB(50 nm)/Al. a**, MPC curves measured at zero bias with excitation photon energies ranging from 2.94 to 4.04 eV. The solid line is fitting results using empirical equation (2), with $MPC_\infty = 1.8$ and $B_{0HF} \sim 140$ mT, respectively. **b**, MPC curves measured at different bias conditions with fixed excitation photon energy at 3.20 eV. The black solid line is fitting result using equation (2), with $MPC_\infty = 1.8$ and $B_{0HF} \sim 140$ mT. The blue solid line is fitting result using equation (3), with $MPC_{HF\infty} = 1.8$, $MPC_{LF\infty} = 2.0$, $B_{0HF} = 140$ mT and $B_{0LF} = 8$ mT, respectively.

Figure 1 shows the *MPC* curves of device $D_{N50}$, ITO/NPB (50 nm)/Al. The *MPC* curves measured at different photon excitation energies, with $E_{hv}$ ranging from 2.97 eV to 4.04 eV, under zero bias condition, are shown in Fig. 1(a). Figure 1(b) illustrates the *V* dependent *MPC* curves with $E_{hv}$ = 3.20 eV, as *V* varying from -0.5 V to 1.4 V. As shown in Fig. 1(a), the *MPC* is positive and does not show significant $E_{hv}$ dependence. The line shape can be fitted by the empirical equation [7],

$$MPC(B) = \frac{MPC_\infty \cdot B^2}{(|B| + B_0)^2} \quad (2)$$

with $MPC_\infty = 1.8$ and $B_0 = 140$ mT, respectively, where $MPC_\infty$ is the *MPC* at an infinite *B* field and $B_0$ is the characteristic field width. The fitting result of $B_0$, ~140 mT, is significantly larger than those reported [7]. We refer this *MPC* behaviour as the high field (HF) component. While with fixed $E_{hv}$ = 3.20 eV, the *V* dependent *MPC* curves can be categorized into two groups, as shown in Fig. 1(b). The line shapes of *MPC* do not show significant bias dependence. For $V < V_t$, where $V_t$ is the turn-on voltage which approximately is 0.9 V for $D_{N50}$, the *MPC* curves can be fitted with the HF component, i.e. using equation (2) with $MPC_\infty = 1.8$ and $B_{0HF} \sim 140$ mT. When $V > V_t$, a new low field (LF) component emerges while the HF component remains unchanged. The *MPC* curves can be fitted by

$$MPC(B) = \frac{MPC_{\infty HF} \cdot B^2}{(|B| + B_{0HF})^2} + \frac{MPC_{\infty LF} \cdot B^2}{B^2 + B_{0LF}^2} \quad (3)$$

where the first term is the HF component and the second term is the LF component, with $MPC_{HF\infty} = 1.8$, $MPC_{LF\infty} = 2.0$, $B_{0HF} = 140$ mT and $B_{0LF} = 8$ mT, respectively. To understand the observed *MPC* effect, detail analyses of the *PC* generation mechanism is necessary and identify the prime magnetic field affects.

In rigid band approximation [23], the highest occupied molecular orbital (HOMO) of NPB lies about 0.6 eV below the ITO Fermi level ($E_F$), whereas the lowest unoccupied molecular orbital (LUMO) lies about 1.8 eV above Al $E_F$ (ref. 24, 25), as shown in Fig. 2(a). Schematic band diagrams of device $D_{N50}$, under different bias conditions, are shown in Fig. 2(b) and (c). The mechanism of *PC* generation includes four basic processes: (i) exciton generation. In OSEs, absorption of a photon can only generate singlet excitons ($S_X$), and triplet excitons ($T_X$) will be generated from $S_X$ through ISC [26,27]. (ii) exciton diffusion. Since excitons are neutral species, their motions are not influenced by electric field and they diffuse via random hops. In principle, $T_X$ has larger diffusion length (> 10 nm) than that of $S_X$ (a few nanometers) [27,28]. And in fact, an increasing number of organic solar sells relies on the diffusion of $T_X$ to dissociating interface [27, and references therein]. (iii) exciton dissociation. Given the typically large exciton binding energy in NPB and/or Alq$_3$, about 1 eV [29], the dissociation of excitons in bulk can be ignored but only occurs at metal-organic interfaces (M/OSE), where an energy compensation, $\Delta$, is available [27,30], as shown in Fig. 2(b) and (c). The number of $S_X/T_X$ diffusing to the M/OSE interfaces actually controls the free carrier generation. The dissociated electrons and holes at M/OSE interface will either annihilate or further separate to become free carriers. The energy compensation $\Delta$ plays important role in exciton dissociation, while the electric field suppresses the annihilation process and benefit the effective dissociation. (iv) carrier transport in OSE. The generated free carriers at M/OSE interface will drift/diffuse, described as polaron hopping from molecular to molecular, towards respective electrodes. The positive/negative polarons will either reach the opposite electrodes or recombines to form PPs, depending on the PP formation rate and the carrier mobility ($\mu$) which determines the carrier transport time through the OSEs.

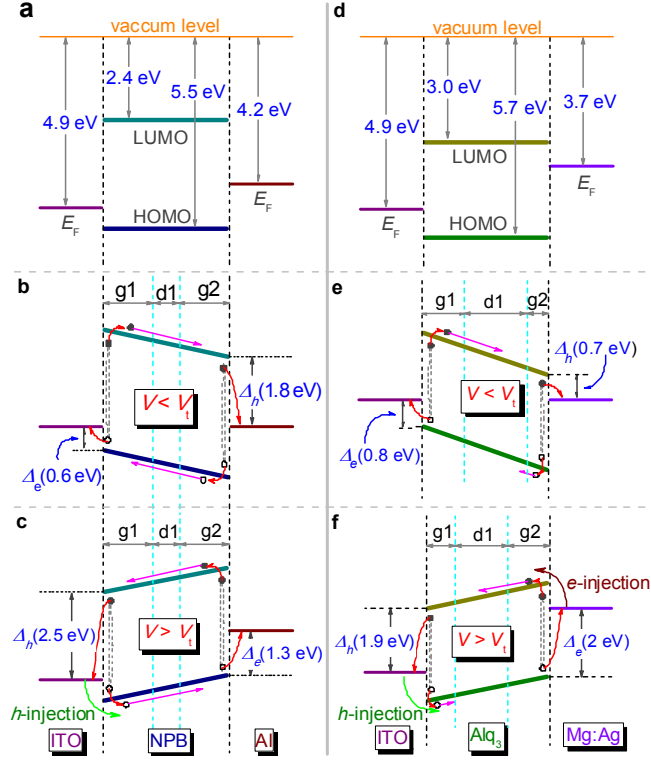

**Figure 2 Schematic band diagrams. a**, The ITO and Al electrodes Fermi levels $E_F$, and LUMO/HOMO energy levels of NPB. **b**, Band diagram of ITO/NPB(50 nm)/Al ($D_{N50}$) when $V < V_t$. **c**, Band diagram of $D_{N50}$ when $V > V_t$. **d**, The ITO and Mg:Ag electrodes Fermi levels $E_F$, and LUMO/HOMO energy levels of Alq$_3$. **e**, Band diagram of ITO/Alq$_3$(50 nm)/Mg:Ag ($D_{A50}$) when $V < V_t$. **f**, Band diagram of $D_{A50}$ when $V > V_t$. $\Delta$ represents energy compensation for the exciton dissociations at M/OSE interfaces. In cases of $V < V_t$ and $V > V_t$, $\Delta$ for releasing electrons and/or holes are different. Region g1/g2 represents the generation zone, in which the excitons can diffuse to M/OSE interface and contribute to dissociation. Region d1 represents the zone that out of the exciton diffusion length from M/OSE interfaces.

From *PC* generation processes described above, it can be seen that any magnetic field induced changes of carrier mobility, and/or the number of excitons that can diffuse to M/OSE interface will lead to photocurrent modulation. In our small molecular OSE based sandwiched structures, we consider two processes, both include carrier spin information that are magnetic field dependent. One is the magnetic field induced decrease of P-T$_X$ quenching (TPQ process) [14,31], and another is the magnetic field induced increasing of ISC rate between singlet and triplet states (ISC process) [26]. TPQ process can be described as

$$P + T_X \leftrightarrow (P, T_X) \xleftrightarrow{k_q} P + S_0 \qquad (4)$$

where $S_0$ is the ground state and $k_q$ is the quenching rate which decreases with increasing magnetic field [14,31]. The left-hand side describes the scattering of polarons by T$_X$, and this scattering is field independent. Changing of T$_X$ density will result in modulation of photocurrent, because it will either affect the scattering of polarons, altering the polaron mobility, or change the number of excitons that can contribute to dissociation at M/OSE interface. The ISC happens in both among exciton states and PP states, depending on the singlet/triplet generation mechanism. In a photon absorption case, due to the fact that photon excitation can only generate S$_X$, the ISC translates S$_X$ into T$_X$. While in a polaron-polaron recombination case, the ISC translates triplet polaron pairs (T$_{PP}$) into singlet polaron pairs (S$_{PP}$) due to their density difference (formation ratio is 3:1) and energy degeneracy. But in both cases, the ISC rate $k_{ISC}$ will increases when an external magnetic field applied.

The $B$ field induced increase of $k_{ISC}$ leads to an increment of T$_X$, consequently inducing a positive *MPC* component by increasing the total number of excitons that can diffuse to M/OSE interface to dissociate, as well as a negative *MPC* component by increasing the P-T$_X$ scattering probability to limit polaron mobility. Both the positive and the negative components follow the same $B$ dependence as $k_{ISC}(B)$. The $B$ field induced decrease of $k_q$, saving more T$_X$, will also induce both positive and negative *MPC* effects, following a same $B$ dependence. Obviously, the signs of *MPC*$_{ISC-X}$ and *MPC*$_{TPQ}$ should be the same because both originate from the increase of T$_X$ density. The PP formation can be considered as a secondary process, and the increase of S$_{PP}$, though $B$ field induced increase of $k_{ISC-PP}$, has a device dependent effect on the *MPC*. When the values of e/h mobility differ a lot and the PP formation zone locates near one M/OSE interface due to carrier accumulation, the $k_{ISC-PP}(B)$ effect should be considered, otherwise, it can be ignored, as will be discussed later.

The values of e/h mobility in NPB are large and in same order of magnitude [24], $\mu_e \sim \mu_h \sim 10^{-3}$ cm$^2$V$^{-1}$s$^{-1}$. Both dissociated electrons and holes will drift towards opposite electrodes. No carrier accumulation exists in this device and $k_{ISC-PP}(B)$ can be ignored here. When $V < V_t$, the P-T$_X$ quenching can be ignored since the density of free polarons is quite low. In Fig. 1(a), the *MPC* curves measured under zero bias is mainly from the $B$ induced T$_X$ increase through $k_{ISC-X}(B)$. The fitting result reveals that $k_{ISC-X}(B)$ is a HF effect following the non-Lorenze line shape. The positive *MPC*$_{ISC-X}$ indicates that $D_{N50}$ can be treated as "generation" limited device. When $V > V_t$, holes are injected into NPB layer and the P-T$_X$ quenching process is activated. The emerged LF component corresponds to the magnetic field effect on $k_q$, which is a low field effect and has a Lorenze line shape. The fitting result, that the values of *MPC*$_{ISC-X}$ and *MPC*$_{TPQ}$ are both positive, is consistent with what we just discussed that both components are originate from the increase of $T_X$ density.

Since OMFEs are universal phenomena in OSEs, the model based on $B$ field induced increase of $k_{ISC}$ and decrease of $k_q$ conjectured from $D_{N50}$, should also apply to other OSE based devices. Thus we studied Alq$_3$ based device ITO/Alq$_3$ (50 nm)/Mg:Ag ($D_{A50}$). Different from NPB, Alq$_3$ is usually used as an electron transporting and light emitting layer. The value of hole mobility is orders of magnitude lower that the electron mobility in Alq$_3$, $\mu_e \sim 10^{-6}$ cm$^2$V$^{-1}$s$^{-2}$ and $\mu_h \sim 10^{-8}$ cm$^2$V$^{-1}$s$^{-1}$ (ref. 24). Because of its low mobility, the dissociated holes will accumulate near the M/OSE interface, and will decrease the exciton dissociation probability due to the Columbic attraction. Therefore, the $PC$ is mainly come from dissociated electrons and its following drift towards opposite electrode. Most of the holes contribute to $PC$ through forming PPs with electrons near M/OSE interface. Since the PP forms near M/OSE interface, the $k_{ISC-PP}(B)$ will affect the $MPC$ significantly. Increase of $k_{ISC-PP}$ will converts more $T_{PP}$ into $S_{PP}$, and effectively decreases the accumulated holes, because the singlet states recombine much more quickly. As a result of the decrease in hole accumulation, the exciton dissociation probability at M/OSE interface increases, resulting a positive $MPC$ component. When $V < V_t$, P-$T_X$ quenching can be ignored due to free polaron density is low. The $MPC$ is a result of both $k_{ISC-PP}(B)$ and $k_{ISC-X}(B)$, with $MPC_{ISC-PP}$ positive while $MPC_{ISC-X}$ having both positive and negative components.

Figure 3(a) shows the $E_{hv}$ dependent $MPC$ curves of $D_{A50}$ measured at zero bias. All the $MPC$ curves consist of a LF positive component and a HF negative component. It is reasonable to assume that both $MPC_{ISC-PP}$ and $MPC_{ISC-X}$ in $D_{A50}$ follow same line shape, a non-Lorenze type as in $D_{N50}$. The solid lines in Fig. 3(a) are fitting results using the following equation:

$$MPC(B) = \frac{MPC_{\infty LF} \cdot B^2}{(|B| + B_{0LF})^2} + \frac{MPC_{\infty HF} \cdot B^2}{(|B| + B_{0HF})^2} \quad (5)$$

with $B_{0LF} = 4.6$ mT and $B_{0HF} = 128$ mT. The fitting results of $MPC_{LF\infty}$ and $MPC_{HF\infty}$ as a function of $E_{hv}$ are shown in Fig. 3(b). Obviously, that the LF and HF components correspond to $MPC_{ISC-PP}$ and $MPC_{ISC-X}$, respectively. The fact, that $k_{ISC-X}(B)$ in Alq$_3$ is a HF effect and follows non-Lorenze line shape, is consistent with that in NPB. The negative value of $MPC_{ISC-X}$ indicates that $D_{A50}$ is "transport" limited at zero bias. As shown in Fig. 3(b), as $E_{hv}$ decreases, $MPC_{ISC-X}$ decreases and $MPC_{ISC-PP}$ increases. This is because that, as $E_{hv}$ decreases, relatively more light was absorbed in d1 and g2 region (Fig. 2(b)), enhancing the P-$T_X$ scattering effect, and consequently the negative component of $MPC_{ISC-X}$. And this absorption region change also enhances the hole accumulation at M/OSE interface, making the $k_{ISC-PP}(B)$ effect more pronounced, increasing the positive $MPC_{ISC-PP}$ value.

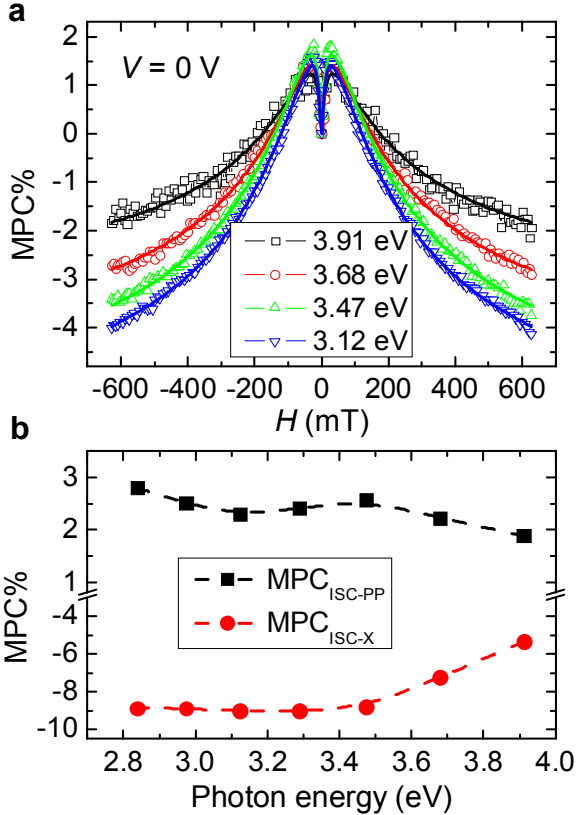

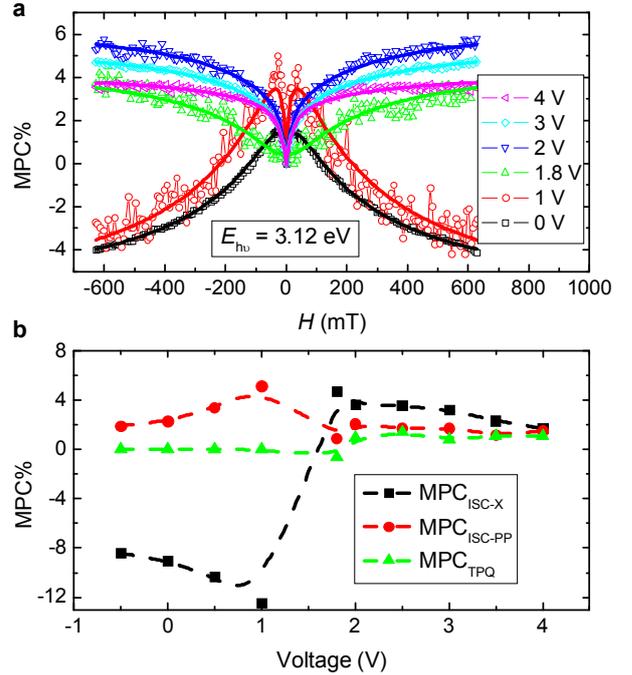

**Figure 3 MPC of device ITO/Alq$_3$(50 nm)/Mg:Ag at zero bias. a**, MPC curves measured with different excitation photon energies. The solid lines are fitting results using equation (5), with $B_{0LF} = 4.6$ mT and $B_{0HF} = 128$ mT, respectively. **b**, The fitting results of $MPC_{LF\infty}$ ($MPC_{ISC-PP}$) and $MPC_{HF\infty}$ ($MPC_{ISC-X}$) as a function of $E_{hv}$.

**Figure 4 MPC of device ITO/Alq$_3$(50 nm)/Mg:Ag with $E_{hv} = $ 3.12 eV. a**, MPC curves measured at different bias conditions indicated. The solid lines are fitting results using equation (6), with $B_{0LF-PP} = 4.6$ mT, $B_{0HF-X} = 128$ mT and $B_{0LF-q} = 27$ mT, respectively. **b**, The fitting results of $MPC_{ISC-PP}$, $MPC_{ISC-X}$ and $MPC_{TPQ}$ as a function of applied bias.

In order to investigate the magnetic field effect on P-$T_X$ quenching process in Alq$_3$, $V$ dependent $MPC$ curves of $D_{A50}$ with $E_{hv} = 3.12$ eV were also measured, and the results are shown in Fig. 4(a). When $V < V_t$, MPC is dominated by LF positive $MPC_{ISC-PP}$ and HF negative $MPC_{ISC-X}$. When $V > V_t$, carriers are injected and the P-$T_X$ quenching is activated. Now the MPC consists of $MPC_{ISC-PP}$, $MPC_{ISC-X}$ and $MPC_{TPQ}$. Based on that both the line shapes and field widths of $MPC_{ISC-PP}$ and $MPC_{ISC-X}$ should remain the same as those at zero bias conditions and the $MPC_{TPQ}$ should have a Lorenze line shape and is a LF component the same as that in NPB, the observed MPC curves can be fitted using the following equation shown as solid lines in Fig. 4(a)

$$MPC(B) = \frac{MPC_{\infty LF-PP} \cdot B^2}{(|B| + B_{0LF-PP})^2} + \frac{MPC_{\infty HF-X} \cdot B^2}{(|B| + B_{0HF-X})^2} + \frac{MPC_{\infty LF-q} \cdot B^2}{B^2 + B_{0LF-q}^2} \quad (6)$$

with $B_{0LF-PP}$ = 4.6 mT, $B_{0HF-X}$ = 128 mT and $B_{0LF-q}$ = 27 mT. The fitting result of $B_{0LF-q}$ = 27 mT, indicating that the P-$T_X$ quenching in Alq$_3$ is also a LF effect, consistent with that in NPB. The fitting values of $MPC_\infty$ for various components as a function of $V$ were shown in Fig. 4(b). The $MPC_{ISC-X}$ changed from negative to positive when the $V$ crosses $V_t$. This is because that the $e/h$ generation region/interface for exciton dissociation interchanged as $V$ across $V_t$, enhancing the generation capability significantly due to energy compensation $\Delta$ increase, as shown in Fig. 2(e) and (f) which illustrate the case of $V < V_t$ and $V > V_t$ respectively. Then the positive component of $MPC_{ISC-X}$ increases and $D_{A50}$ cannot be treated as "transport" limited, resulting in the sign change.

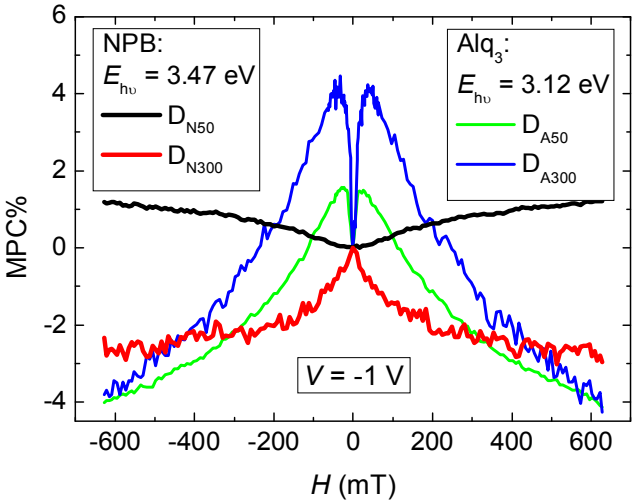

**Figure 5 Comparison of MPC curves of devices with OSE layer thickness of 50 nm and 300 nm.** For NPB based devices, $E_{h\nu}$ is 3.47 eV and for Alq$_3$ based devices is 3.12 eV. For devices with 300 nm thick OSE layer, MPC curves were measured under a -1 V bias condition in order to reach a uniform electric field.

We have shown that although the measured MPC results are quite different between NPB and Alq$_3$ based devices, the underline magnetic field dependent physical processes are the same. The reason of the divergence between $D_{N50}$ and $D_{A50}$ is the difference in their device physics, such as energy bands alignment and carrier transport behaviors. In $D_{A50}$, it has been shown in "transport" limited case, $MPC_{ISC-X}$ is negative. To further verify our model, we fabricates a NPB based "transport" limited device, by simply increase the thickness of NPB layer to 300 nm ($D_{N300}$). Together an Alq$_3$ based device with the same thickness ($D_{A300}$) was also fabricated. It is expected from our analysis that in this NPB "transport" limited device, negative $MPC_{ISC-X}$ should be observed, while in device $D_{A300}$, the signature of the line shape should be similar with that of $D_{A50}$. Figure 5 shows the MPC curves of $D_{N300}$ and $D_{A300}$ with $E_{h\nu}$ = 3.47 eV and $E_{h\nu}$ = 3.12 eV, respectively. Both were measured under $V$ = -1 V bias condition, in order to achieve a uniform electric field. As can be seen in Fig. 5, the MPC value of $D_{N300}$ is negative and dominated by a HF component, while the MPC curve of $D_{A300}$ has similar line features with that of $D_{A50}$. This further confirms that the OMFEs originate from the magnetic field induced increase of $k_{ISC}$ and decrease of $k_q$ and the device physics controls the sign change and the observed field dependence.

In conclusion, we have identified the underlying magnetic field dependent physical processes that are responsible for the observed OMFEs. The ISC between singlet and triplet states is enhanced by an external magnetic field. It is revealed that the ISC between excitons is a high field effect, while that between polaron pairs is a low field effect. The P-$T_X$ quenching is a low field effect and the quenching rate is reduced by an external magnetic field. It has been shown that the divergence of observed OMFEs, such as sign changes and operating conditions dependences, originates from the changes in device physics. The result, that ISC-X is a high field effect and ISC-PP is a low field effect, reveals that magnetic field dependence of ISC is inversely proportional to the energy difference between singlet and triplet states. Although both can be treated as low field effect, the characteristic field width of TPQ process in Alq$_3$, ~27 mT, is apparently larger than that in NPB, ~8 mT. This implies that electron-$T_X$ quenching and hole-$T_X$ quenching have different magnetic field dependences. We believe that our result is benefit to further theoretical study of microscopic mechanism of OMFEs.

**Methods:**

The 3×3 mm devices were fabricated on ITO patterned glass substrates. Before transferred into a vacuum chamber, the substrates were rigorously cleaned following by UV o-zone treatment. The organic layer and cathode electrodes were evaporated sequentially in a standard vacuum chamber with a base vacuum of 10$^{-6}$ Torr. After fabrication the sandwiched structures, the devices were transferred to an inert N$_2$ atmosphere glove box and encapsulated. A Xe-lamp together with a Spex1681 spectrometer was used as the excitation source, and a standard lock-in technique was adopted to measure the photocurrent. The light was incident through the glass substrate and ITO. A HP4155A semiconductor parameter analyzer was used to measure the device current-voltage characteristics and as a constant voltage source. An electromagnet with field range from 0 to 800 mT was used to apply the magnetic field. All the measurements were conducted at room temperature. No magnetic field direction dependence was observed and the magnetic field was applied perpendicularly to the device current direction.

**Acknowledgements**

This work was supported by the Research Grant Council of Hong Kong SAR government via grant number HKUST16/CRF/08 and HKU10/CRF/08.